\documentclass[twocolumn,prl,showpacs,superscriptaddress,preprintnumbers,amsmath,amssymb]{revtex4}

\usepackage{graphicx}
\usepackage{color}
\begin{document}


\title{Direct observation of domain-wall configurations transformed by spin currents}
\author{M. Kl{\"a}ui}

\affiliation{Fachbereich Physik, Universit\"at Konstanz, Universit\"atsstr. 10, D-78457 Konstanz,
Germany}

\affiliation{IBM Research, Zurich Research Laboratory, CH-8803 R\"uschlikon, Switzerland}

\author{P.-O. Jubert}
\thanks{Corresponding author; Electronic address: pju@zurich.ibm.com}
\author{R. Allenspach}
\author{A. Bischof}

\affiliation{IBM Research, Zurich Research Laboratory, CH-8803 R\"uschlikon, Switzerland}

\author{J. A. C. Bland}
\affiliation{Cavendish Laboratory, University of Cambridge, Madingley Road, Cambridge, CB3 0HE,
U.K.}

\author{G. Faini}
\affiliation{Laboratoire de Photonique et de Nanostructures-CNRS,
Route de Nozay, F-91460 Marcoussis, France}

\author{U. R{\"u}diger}
\affiliation{Fachbereich Physik, Universit\"at Konstanz,
Universit\"atsstr. 10, D-78457 Konstanz, Germany}
\author{C. A. F. Vaz}
\affiliation{Cavendish Laboratory, University of Cambridge,
Madingley Road, Cambridge, CB3 0HE, U.K.}

\author{L. Vila}
\affiliation{Laboratoire de Photonique et de Nanostructures-CNRS,
Route de Nozay, F-91460 Marcoussis, France}

\author{C. Vouille}
\thanks{Permanent address: Laboratoire de Physique des Solides, Universit\'e
Paris-Sud, F-91405 Orsay, France.} \affiliation{Laboratoire de Photonique et de
Nanostructures-CNRS, Route de Nozay, F-91460 Marcoussis, France}

\begin{abstract}
Direct observations of current-induced domain-wall propagation by spin-polarized scanning electron
microscopy are reported. Current pulses move head-to-head as well as tail-to-tail walls in
sub-micrometer Fe$_{20}$Ni$_{80}$ wires in the direction of the electron flow, and a decay of the
wall velocity with the number of injected current pulses is observed. High-resolution images of
the domain walls reveal that the wall spin structure is transformed from a vortex to a transverse
configuration with subsequent pulse injections. The change in spin structure is directly
correlated with the decay of the velocity.
\end{abstract}

\maketitle
\newpage

New approaches to the switching of magnetic nanostructures are currently being investigated
intensively because easy and reproducible switching is critical to the use of any spintronic
device. Beyond conventional switching by magnetic fields, a promising approach is current-induced
magnetization switching, which was shown to be able to reverse the soft layer of a
giant-magnetoresistive multi-layer structure~\cite{cims}. As recently demonstrated, spin-transfer
effects can also be used to displace a magnetic domain wall by injecting current
\cite{GBC+03,TFP03,VAA+03,YON+04,KVBWF+03,klaeuisubmitted}. This effect shows potential for novel
memory and logic devices based on domain-wall propagation \cite{AXC+02} as it could simplify
designs by eliminating magnetic field-generating circuits. While field-induced domain-wall motion
is well established, current-induced domain-wall motion is still not thoroughly understood.
Several effects occur when large electrical currents flow across a domain wall, the most prominent
ones being the action of the field created by the current itself (the so-called Oersted field) and
the spin momentum transfer, also known as spin torque effect~\cite{Berger}. Domain drag is
believed to be important only in thick films \cite{Berger74}, and linear momentum transfer only at
high frequencies or for very narrow domain walls \cite{Saitoh, TataraPRL2004}.

The understanding of the spin torque effect has been extended recently by various approaches that
treat the interactions between the spin current and the magnetization, but the appropriate form of
the spin-transfer contribution still is the subject of much debate. Most theoretical models
describing the current interaction with wide domain walls are based on the adiabatic
approximation, in which the spin polarization of the current is assumed to remain aligned with the
magnetization vector in the domain wall \cite{Slonczewski96, TataraPRL2004, LiPRL2004, LiPRB2004,
ThiavilleJAP2004}. These models explain current-induced wall motion qualitatively, but only for
currents much larger than observed experimentally \cite{LiPRB2004,ThiavilleJAP2004}. Corrections
to the adiabatic approximation have been introduced
\cite{ZhangCM2004,ThiavilleCM2004,WaintalEPL2004}, with an additional nonadiabatic term related to
the spatial mistracking of spins between conduction electrons and local magnetization. While some
of these approaches predict a wall motion at reduced current density \cite{ThiavilleCM2004} and
some find wall velocities of the order of magnitude observed experimentally \cite{ZhangCM2004},
the parameters and the results of the calculations vary significantly. Interestingly, all theories
predict that the spin current modifies the wall structure, but they disagree on whether this
change is transient or permanent, and whether it is a subtle distortion or even a change of wall
type. Thus observing domain-wall spin-structure changes is expected to provide important input to
refine current theories.


Experimentally, the domain-wall displacement, the velocities and the critical current densities
have recently been measured in various single-layer geometries \cite{VAA+03,
YON+04,KVBWF+03,TFP03} and in multilayer wires \cite{GBC+03}. Interestingly, it was found that the
walls do not always move with constant velocity or even stop moving \cite{YON+04,klaeuisubmitted}.
This has been attributed mainly to extrinsic mechanisms, such as materials degradation or pinning.
Alternatively, it has been suggested that an intrinsic magnetic effect, such as a change in spin
structure, could play a role \cite{klaeuisubmitted}. To our knowledge, experiments to test this
conjecture have not yet been done.

In this letter we report current-induced domain-wall displacement experiments that are combined
with \emph{in-situ} high-resolution magnetic imaging. Effects of current pulses on head-to-head
domain walls in straight sub-micrometer Fe$_{20}$Ni$_{80}$ (Permalloy) wires are imaged using
spin-polarized scanning electron microscopy (spin-SEM or SEMPA). Variations of the domain-wall
velocity with the number of current pulse injections at a constant current density are compared
and correlated with modifications of the nanoscale domain-wall configuration induced by the
current.

We investigate Fe$_{20}$Ni$_{80}$ wires with a zigzag geometry, see Fig. 1(a). Straight wire
segments 20 $\mu$m long are connected by bends that consist of 45$^{\circ}$ ring sections having a
radius of 2 $\mu$m. We have fabricated wires with widths ranging from 100 nm to 500 nm and
thicknesses from 6 nm to 27 nm on a Si substrate covered by native oxide using electron-beam
lithography and a two-step lift-off process as described in Ref. \onlinecite{KVB+02}.
Fe$_{20}$Ni$_{80}$ was deposited by molecular beam epitaxy at $\sim 5 \times 10^{-10}$ mbar,
followed by a thin 1.5 nm Fe layer, a 2 nm Au capping layer to prevent oxidation, and subsequent
lift-off. The Fe layer enhances the magnetic contrast during imaging without altering the magnetic
properties significantly. Finally, 100 nm thick sputtered Au contacts are defined in a second
lithography step to contact each wire individually.

The current injection experiments and magnetic imaging of both in-plane magnetization components
were performed in our spin-SEM setup \cite{Allenspach1994JMMM}. Topography and magnetization
distribution are determined simultaneously and with a lateral resolution of $\simeq 20$ nm.  Prior
to imaging, the Au capping layer was removed by mild Ne$^+$ ion bombardment.

\begin{figure}[t]
  \begin{center}
  \leavevmode
  \includegraphics[width=80mm]{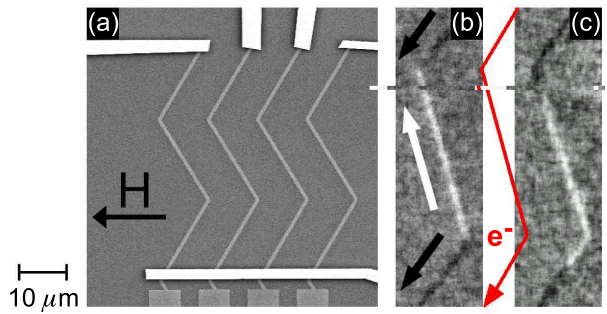}
  \end{center}
  \vspace*{-5mm}
  \caption{(color online) (a) Topographic image of the device structure showing the Au contacts
(white) and the four zigzag Fe$_{20}$Ni$_{80}$ wires (light grey) with square pads at the bottom.
(b) Magnetization configuration in a wire after magnetizing with a field pulse along the direction
indicated by the arrow. White (black) corresponds to the magnetization pointing up (down) within
the plane; a head-to-head wall is formed at the top bend, a tail-to-tail wall at the bottom. (c)
After injection of a single 10 $\mu$s long current pulse through this wire, both domain walls have
moved in the direction of the electron flow as indicated by the arrow.}
    \label{FigSpinSEMImages}
\end{figure}

The zigzag geometry is chosen as it allows the magnetic configuration of the wires to be
controlled by application of an external magnetic field. After saturation along the direction
indicated in Fig. 1(a) and relaxation of the field to zero, shape anisotropy forces the
magnetization to form domains of alternating directions in adjacent segments, see Fig. 1(b). At
the bends head-to-head and tail-to-tail walls form \cite{TNNY99}. The dimensions of the wire
control the type of the domain walls \cite{McMichael,klaeuiaplinprint}. In this paper we
concentrate on 500 nm wide and 10 nm thick wires that result in vortex walls.

\begin{figure}[b]
  \begin{center}
  \leavevmode
  \includegraphics[width=80mm]{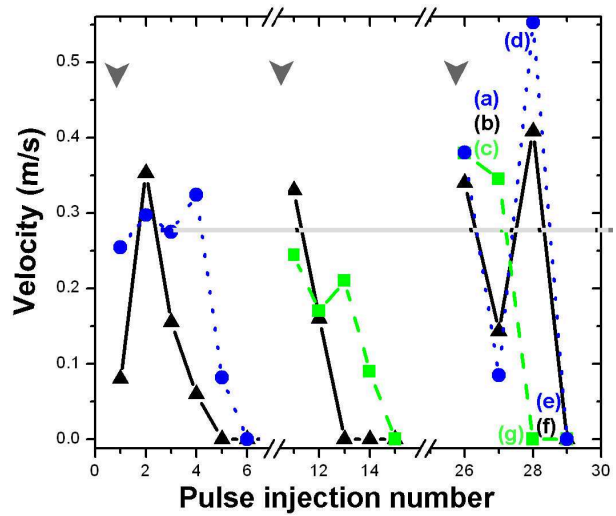}
  \end{center}
  \vspace*{-5mm}
  \caption{(color online) Domain-wall velocity as a function of pulse injection number determined
from spin-SEM images (wall $\alpha$: blue circles, dotted line; $\beta$: black triangles, solid
line; $\gamma$: green squares, dashed line). The magnetic state has been re-initialized by a
magnetic field before pulses 1, 11, and 26, as indicated by the arrows. After pulses 26 to 28,
high resolution images of the domain wall have been taken. The labels are related to the images
shown in Fig. 3. Statistical uncertainty of the wall velocity is 0.05 m/s.}
    \label{FigVelocity}
\end{figure}

After initializing the system with a magnetic field $\geq60$ kA/m, a head-to-head domain wall is
located at the upper bend and a tail-to-tail wall at the lower bend. Then a single current pulse
of 10 $\mu$s duration is injected with a current density of $2.2 \times 10^{12}$ A/m$^2$. This
current density is 10\% higher than the threshold current density at which domain-wall motion sets
in, which was measured to be the same for walls located at a bend or in the straight part of the
wire within an accuracy of 10\%. After injection, both walls have moved in the direction of the
electron flow, see Fig. 1(c). The distances the head-to-head and tail-to-tail walls have traveled
are 3.0 $\mu$m and 2.9 $\mu$m, respectively, which yields a mean wall velocity of 0.3 m/s for the
10 $\mu$s pulse. As both walls propagate in the same direction, the Oersted field can be excluded
as a possible cause for wall motion: Our observation is consistent with an explanation based on
the spin-torque effect due to the current pulse. Correspondingly, injecting a current pulse with
opposite polarity moves both walls back to the bends.

\begin{figure*}[t]
  \begin{center}
  \leavevmode
  \includegraphics[width=160mm]{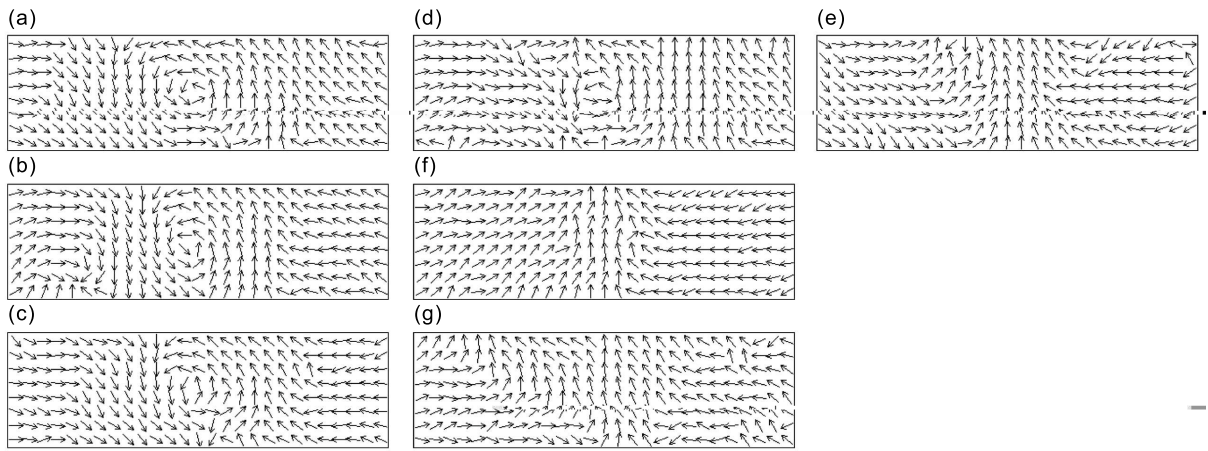}
  \end{center}
  \vspace*{-5mm}
  \caption{High resolution experimental images of the spin structure of domain walls $\alpha$:
(a,d,e), $\beta$: (b,f) and $\gamma$: (c,g). After the first current injection 26, the walls are
all of vortex type (a,b,c). After injection 27, wall $\gamma$ has stopped moving and undergone a
drastic transformation to a very distorted transverse wall type (g), whereas the mobile wall
$\alpha$ has a vortex core and a large transverse component (d). After injection 28, walls
$\alpha$ and $\beta$ have also stopped and changed to transverse walls (e,f). The arrow images are
constructed from the two orthogonal in-plane magnetization components taken by spin-SEM. Image
size: 1600 nm by 500 nm.}
    \label{FigWallImages}
\end{figure*}

To exclude effects related to the curved geometry at the bend, we consider in the following wall
propagation in the straight part of the wire, i.e., where the wall is located after the first
current-pulse injection. Starting from this configuration, current pulses (10 $\mu$s duration with
current density $2.2 \times 10^{12}$ A/m$^2$) are injected and the domain wall velocity for each
pulse is determined. Figure 2 shows the evolution of the velocity with the number of injected
current pulses for three different walls. After initialization, the walls propagate under pulse
injection. After a few injections, however, the walls stop moving. The starting velocity can be
retrieved by re-initializing the sample with a magnetic field as described above, which was
carried out before injections 11 and 26. The complete stopping of the walls was a general
observation for all walls in our straight wire segments. The number of injections after which the
wall stops moving varies from a few to a few tens. We note that wall motion in general is a
stochastic process, and non-constant wall velocities have also been observed in other experiments
\cite{YON+04,klaeuisubmitted}.

To understand the wall-velocity decay, we have taken high-resolution images of the spin structure
of three domain walls (labeled $\alpha, \beta, \gamma$) after subsequent injections (26 to 29), as
shown in Fig. 3. The first pulse (injection 26) moves the domain walls into the straight part of
the wire, similar to Fig. 1(c). All three walls are vortex walls with a well centered core and a
width $w$ ranging from 400 nm to 660 nm, as determined from a fit with the usual tanh$(x / w)$
function (Fig. 3(a)-(c)). A micromagnetic simulation of a relaxed vortex wall in a perfect wire
reproduce this spin structure with $w = 400$ nm. The next injection modifies the structure of wall
$\alpha$ (Fig. 3(d)): While the wall still contains a vortex core, it has acquired a transverse
component. The subsequent injection, 28, drastically changes the structure of the wall (Fig.
3(e)): The vortex is eliminated, and a narrow (210 nm) distorted transverse wall has formed.
Further injections do not move the wall anymore.

The other two walls display the same behavior: Wall $\beta$, starting from a vortex (Fig. 3(b)),
has attained a transverse structure after injection 28 that is very similar to that of wall
$\alpha$ (Fig. 3(f)). Likewise, it does not move anymore with subsequent injections. Wall $\gamma$
already fails to move after injection 27. Again, the wall has a strongly distorted transverse
character, with the vortex core annihilated or expelled from the structure (Fig. 3(g)).

Thus, in all three cases the walls move as long as they are vortex walls but stop moving when they
attain a transverse structure. From these observations we conclude that a direct correlation
between the spin structure and the domain-wall velocity exists, which we propose to be the cause
for the behavior of the wall velocity observed in our experiments as well as that of others
\cite{YON+04,klaeuisubmitted}.

Defects cannot directly account for the domain walls stopping after a few injections: The walls
have been moved by the current pulses over the entire area between the bends, and have even passed
the position at which they eventually stop a number of times. Moreover, after every
reinitialization and current injection, the walls stop at a different position of the wire.
High-resolution imaging of the different wire sections at which the walls stop does not reveal any
obvious structural defects that might lead to pinning. We can also exclude structural damage to
the material due to the high current densities as a cause for the wall stopping. As seen in Fig.
2, the wall velocity starts with similar values after each re-initialization. In addition, the
resistance of the wires stayed constant at 5 k$\Omega$ over the course of the experiment, which
means that no detrimental effects such as electromigration or excessive heating were discernible.
Hence we conclude that the electrical current induces both motion and distortion of the wall.

Recent theories qualitatively predict some domain-wall distortion induced by the spin current
\cite{LiPRB2004, ZhangCM2004, ThiavilleCM2004, WaintalEPL2004}. For a 1-D N{\'e}el wall, Li and
Zhang predict a transient distortion which builds up during the first few nanoseconds
\cite{LiPRB2004}. Waintal and Viret \cite{WaintalEPL2004} anticipate significant distortions of
the wall structure up to the point at which the wall switches between different types. A step
beyond the 1-D models has been taken by Thiaville et al. \cite{ThiavilleCM2004} with a 2-D
micromagnetic simulation. For a wire narrower than ours they find a periodic transformation of the
wall structure from vortex to transverse, albeit at larger current densities.

While the domain wall motion is caused by the spin torque, the origin of the wall transformation
is less obvious. The most prominent signature of our observation is the breaking of the wall
symmetry. A priori, spin torque alone is not necessarily very effective in achieving this. Only at
current densities much larger than our experimental value do Thiaville et al. report such a
transformation of wall types~\cite{ThiavilleCM2004}. The Lorentz force also breaks the symmetry.
It leads to domain drag in thick films \cite{Berger74}. In our thin films, it is not the dominant
effect for domain wall propagation, but it exerts a transverse force on the perpendicularly
magnetized vortex core. This could help pushing it off the center and eventually expel it from the
wire. Thus while domain wall spin structure modifications and even transitions from vortex to
transverse walls due to spin currents have been predicted, other intrinsic magnetic effects could
play a role. Calculations will have to be carried out for our geometry to discriminate between the
possible explanations and gain a deeper understanding of our observations.

Further to the observation of domain wall transformation, our experiments demonstrate a direct
correlation between the change of wall structure and the reduction of wall velocity. Additional
measurements on wires with different dimensions show that the velocity after field initialization
also depends on the wire width and thickness and hence on the wall width for constant current
density: The velocity is 0.3 m/s for a width of 500 nm and thickness of 10 nm, but 1.2 m/s for a
width of 200 nm and thickness of 27 nm. A detailed systematic study is beyond the scope of this
paper and will be published elsewhere.

Our observations of varying velocities are in striking disagreement with theoretical models
\cite{ThiavilleCM2004, ZhangCM2004, LiPRL2004}, which predict the velocities to be only dependent
on material parameters and on the current density, but not on the type of the wall and its spin
structure. Moreover, our experimental mean velocities are smaller than those calculated by at
least one order of magnitude \cite{ThiavilleJAP2004, ThiavilleCM2004, ZhangCM2004}.

These discrepancies between experiment and calculation are unresolved. We feel that thermal
excitations may play a significant role. At finite temperature, spin waves reduce the spin
polarization of the current that exerts the spin torque on the wall \cite{YamaguchiAPL2005}.
Further theoretical work is needed to quantum mechanically calculate the consequences for the spin
wave dispersion, but also to include finite temperature effects in micromagnetic simulations.

In conclusion we have observed current-induced domain-wall propagation by spin-polarized scanning
electron microscopy. Head-to-head as well as tail-to-tail domain walls in 500 nm wide and 10 nm
thick Fe$_{20}$Ni$_{80}$ wires both move in the direction of the electron flow with a mean
velocity of 0.3 m/s, which is consistent with an explanation based on the spin-torque effect. The
velocity varies, and after a number of pulse injections the walls eventually stop moving. The
original velocity is re-established by re-initializing the sample with a magnetic field.
High-resolution images of the wall structure after consecutive pulse injections show a
transformation from a vortex wall to a distorted transverse wall due to the current. The change in
wall velocity is correlated with a change in the domain-wall spin structure. These results are
largely not reproduced using the theoretical models currently available. Our observation of a
drastic change in wall structure by current is a salient feature, which should stimulate further
development of theory and lead to a deeper insight into the interactions between current and
magnetic domain walls.

This work was partially funded by the CMI Magnetoelectronic Devices project and the 'Deutsche
Forschungsgemeinschaft' (SFB 513). M.K. acknowledges the 'Deutscher Akademischer Austauschdienst'
(DAAD) for financial support.

\bibliographystyle{apsrev}

\end{document}